\documentclass[aps,prc,twocolumn,superscriptaddress,showpacs]{revtex4}
\usepackage{graphicx}

\begin{document}


\title{Alpha-decay branching ratios of near-threshold states in $^{19}$Ne and the astrophysical rate of $^{15}$O($\alpha$,$\gamma$)$^{19}$Ne}


\author{B. Davids}
\email[]{davids@kvi.nl}
\author{A. M. van den Berg}
\author{P. Dendooven}
\author{F. Fleurot}
\altaffiliation{Present address: Department of Physics, Laurentian University, Sudbury ON, P3E 2C6, Canada}
\author{M. Hunyadi}
\author{M. A. de Huu}
\affiliation{Kernfysisch Versneller Instituut, Zernikelaan 25, 9747 AA Groningen, The Netherlands}
\author{K. E. Rehm}
\affiliation{Physics Division, Argonne National Laboratory, Argonne IL 60439}
\author{R. E. Segel}
\affiliation{Department of Physics, Northwestern University, Evanston IL 60208}
\author{R. H. Siemssen}
\author{H. W. Wilschut}
\author{H. J. W\"{o}rtche}
\affiliation{Kernfysisch Versneller Instituut, Zernikelaan 25, 9747 AA Groningen, The Netherlands}
\author{A. H. Wuosmaa}
\affiliation{Physics Division, Argonne National Laboratory, Argonne IL 60439}


\date{\today}

\begin{abstract}
The $^{15}$O($\alpha$,$\gamma$)$^{19}$Ne reaction is one of two routes for breakout from the hot CNO cycles into the $rp$ process in accreting neutron stars. Its astrophysical rate depends critically on the decay properties of excited states in $^{19}$Ne lying just above the $^{15}$O + $\alpha$ threshold. We have measured the $\alpha$-decay branching ratios for these states using the $p(^{21}$Ne,$t)^{19}$Ne reaction at 43 MeV/u. Combining our measurements with previous determinations of the radiative widths of these states, we conclude that no significant breakout from the hot CNO cycle into the $rp$ process in novae is possible via $^{15}$O($\alpha$,$\gamma$)$^{19}$Ne, assuming current models accurately represent their temperature and density conditions.
\end{abstract}

\pacs{26.30.+k, 25.60.Je,  26.50.+x,  27.20.+n}

\maketitle

Novae are thermonuclear runaways initiated by the accretion of hydrogen- and helium-rich material from stellar companions onto the surfaces of white dwarfs in binary systems. Energy production and nucleosynthesis in the hottest novae are determined principally by the CNO, NeNa, and MgAl cycles \cite{gehrz98}. Under high temperature and density conditions, e.g. in accreting neutron stars, breakout from the hot CNO cycles into the $rp$ process occurs \cite{wallace81}, dramatically increasing the luminosity of outbursts and synthesizing nuclei up to masses of 100 u \cite{schatz01}. Several reactions have been suggested as pathways for this breakout \cite{wiescher99}, but only two are currently thought to be possibilities: $^{15}$O($\alpha$,$\gamma$)$^{19}$Ne and $^{18}$Ne($\alpha$,$p$)$^{21}$Na. In astrophysical environments the $^{15}$O($\alpha$,$\gamma$)$^{19}$Ne reaction proceeds predominantly through resonances lying just above the $^{15}$O + $\alpha$ threshold at 3.529 MeV in $^{19}$Ne, as the direct capture component is very small by comparison \cite{langanke86,dufour00}. The reaction rate in novae is determined by the $\alpha$-width $\Gamma_{\alpha}$ of the 4.033 MeV 3/2$^+$ state, owing both to its close proximity to the $^{15}$O + $\alpha$ threshold and its low centrifugal barrier to $\alpha$-capture.

A previous attempt to determine $\Gamma_{\alpha}$ for this state was based upon measurements of $\alpha$-transfer reactions to the analog state in the mirror nucleus $^{19}$F \cite{mao96}. Such determinations, however, are subject to large uncertainties \cite{oliveira97}. Direct measurements of the low energy cross section, which require high-intensity radioactive $^{15}$O beams, are planned. At $^{19}$Ne excitation energies relevant to novae and accreting neutron stars, only the $\alpha$- and $\gamma$-decay channels are open, as the proton and neutron separation energies are 6.4 and 11.6 MeV \cite{audi95} respectively. Hence, by populating these states and observing the subsequent $\alpha$-  and $\gamma$-decays, one can deduce the branching ratio B$_{\alpha}\equiv\Gamma_{\alpha}/\Gamma$. If $\Gamma_{\gamma}$ is also known, one can then calculate $\Gamma_{\alpha}$ and thereby the contribution of each state to the resonant rate of $^{15}$O($\alpha$,$\gamma$)$^{19}$Ne. A pioneering effort of this kind was made by detecting $\alpha$ particles from the decay of $^{19}$Ne states populated via the $^{19}$F($^3$He,$t)^{19}$Ne reaction \cite{magnus90}, but the sensitivity of the experiment was insufficient to measure B$_{\alpha}$ for the critical 4.033 MeV state, which was expected to be of order 10$^{-4}$. Despite vigorous efforts worldwide \cite{kubono02,wiescher02}, up to now no experiment has reached this level.

In an experiment at the Kernfysisch Versneller Instituut, we have obtained branching ratio data of high sensitivity by applying a novel method introduced at Argonne National Laboratory using a different reaction \cite{rehm00}. Populating the important states via the $^{21}$Ne($p,t)^{19}$Ne reaction in inverse kinematics with a $^{21}$Ne beam energy of 43 MeV/u, we detected either $^{19}$Ne recoils or their $^{15}$O $\alpha$-decay products in coincidence with tritons in the Big-Bite Spectrometer (BBS) \cite{berg95}. The large momentum acceptance of the BBS ($\Delta$p/p~=~19\%) allowed detection of either $^{19}$Ne recoils or $^{15}$O decay products along with tritons emitted backward in the center of mass system. Positioning the BBS at 0$^{\circ}$ maximized the yield to the 4.033 MeV 3/2$^+$ state in $^{19}$Ne. This state, whose dominant shell-model configuration is $(sd)^5 (1p)^{-2}$ \cite{fortune78}, was selectively populated by an $\ell$ = 0, two-neutron transfer from the 3/2$^+$ ground state of $^{21}$Ne. Position measurements in two vertical drift chambers (VDCs) \cite{woertche01} allowed reconstruction of the triton trajectories. Excitation energies of the $^{19}$Ne residues were determined from the kinetic energies and scattering angles of the triton ejectiles. The $\gamma$-decays of states in $^{19}$Ne were observed as $^{19}$Ne-triton coincidences in the BBS, whereas $\alpha$-decays were identified from $^{15}$O-triton coincidences.

Recoils and decay products were detected and stopped just in front of the VDCs by fast-plastic/slow-plastic phoswich detectors \cite{leegte92} that provided energy loss and total energy information. A separate array of phoswich detectors was used to identify tritons after they passed through the VDCs. Timing relative to the cyclotron radio frequency signal was also employed for unambiguous particle identification. An Al plate prevented many of the heavy ions copiously produced by projectile fragmentation reactions of the $^{21}$Ne beam in the (CH$_2$)$_n$ target from reaching the VDCs. The spatial extent of the heavy-ion phoswich array was sufficient to guarantee 100\% geometric efficiency for detection of $^{19}$Ne recoils and $^{15}$O decay products for $^{19}$Ne excitation energies~$\leq$~5.5 MeV. This resulted largely from the forward focusing of the $^{19}$Ne recoils, which emerged at angles~$\leq$~0.36$^{\circ}$ for tritons with scattering angles~$\leq$~4$^{\circ}$. The low decay energies of the states studied limited the angular and energy spreads of the $^{15}$O decay products. High geometric efficiency, combined with the excellent background rejection afforded by detecting $^{15}$O nuclei instead of $\alpha$ particles, provided sensitivity to very small $\alpha$-decay branches. 

\begin{figure}\includegraphics[width=\linewidth]{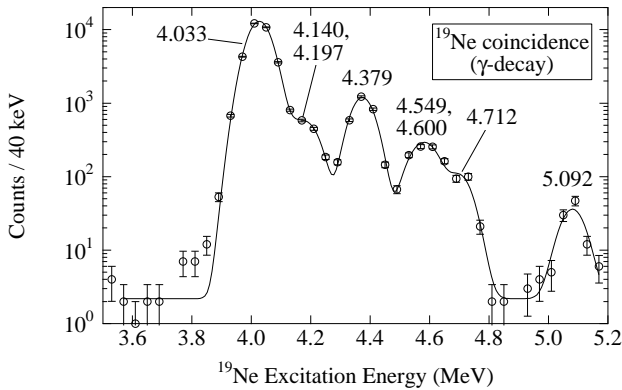} \caption{$^{19}$Ne-triton
coincidences ($\gamma$-decays of states in $^{19}$Ne). The curve is the sum of a constant background and 8 Gaussians centered at the energies of known states in $^{19}$Ne \protect\cite{tilley95}, the widths of which were determined by the experimental resolution of 90 keV FWHM.} \label{fig1} \end{figure}

The $^{19}$Ne excitation energy spectrum obtained from $^{19}$Ne-triton coincidences, representing $\gamma$-decays of states in $^{19}$Ne, is shown in Fig.\ \ref{fig1}. Its most prominent peak is due to the 4.033 MeV 3/2$^+$ state. The 4.379 MeV 7/2$^+$ state is the second most strongly populated. Contributions from other known states \cite{tilley95} are indicated. The experimental resolution of 90 keV FWHM is insufficient to resolve the 4.140 and 4.197 MeV states from one another; the 4.549 and 4.600 MeV states are also unresolved. However, the astrophysically important 4.033 and 4.379 MeV states are well separated from the others. The curve shown in Fig.\ \ref{fig1} is the sum of a constant background and 8 Gaussians centered at the known energies of the states, with standard deviations fixed by the experimental resolution of 90 keV FWHM.

\begin{figure}\includegraphics[width=\linewidth]{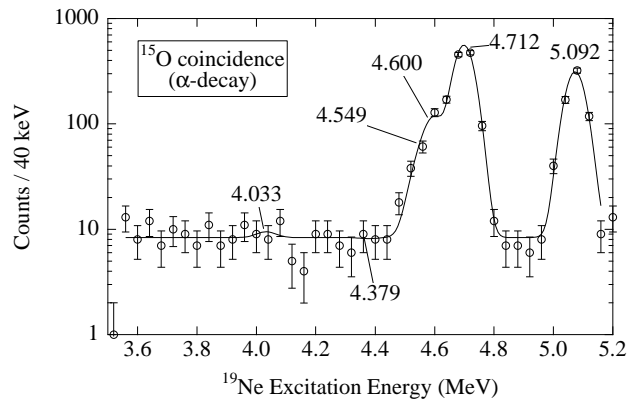} \caption{$^{15}$O-triton
coincidences ($\alpha$-decays of states in $^{19}$Ne). The curve is the sum of a constant background and 6 Gaussians corresponding to known states in $^{19}$Ne, the widths of which were fixed by the experimental resolution. The $^{15}$O + $\alpha$ threshold lies at 3.529 MeV.} \label{fig2} \end{figure}

Fig.\ \ref{fig2} shows the $^{19}$Ne excitation energy spectrum obtained from $^{15}$O-triton coincidences, corresponding to $\alpha$-decays of states in $^{19}$Ne. The observed states are labeled by their energies; the 4.549 and 4.600 MeV states cannot be resolved. A fit consisting of Gaussians plus a constant background is shown as well. The states below 4.549 MeV decay overwhelmingly by $\gamma$-emission, while the higher-lying states observed here decay preferentially by $\alpha$-emission, simply because the larger available decay energy enhances the barrier penetrability. The background represents a larger fraction of the total events in the $^{15}$O-triton coincidence spectrum than in the $^{19}$Ne-triton coincidence spectrum, though it is still very small.  This background is well reproduced by a constant, the value of which is determined to a precision of 7\% (1$\sigma$).

No statistically significant evidence for $\alpha$-decays from the states at 4.033 and 4.379 MeV was observed. For these states the $\alpha$- and $\gamma$-decay spectra were numerically integrated in 100 keV intervals centered at the known energies of the states. These data were  subjected to both Bayesian and classical statistical analyses to determine upper limits on the $\alpha$-decay branching ratios at various confidence levels. The two analyses agreed rather well, with the calculated upper limits differing by less than 20\% in all cases. The Bayesian analysis was found to be more conservative, and has been adopted here. As expected, there is no indication in the data of $\alpha$-decays from the 4.140 and 4.197 MeV states because these decays are hindered by $\ell=4$ centrifugal barriers and low decay energies. For the states at 4.549, 4.600, 4.712, and 5.092 MeV, B$_{\alpha}$ was obtained from the fits described above. The branching ratios are shown in Table\ \ref{table} along with the results of Refs.\ \cite{magnus90,laird01}. Uncertainties in the present branching ratio determinations are purely statistical. Our 90\% confidence level upper limit on B$_{\alpha}$ for the 4.379 MeV state is a factor of 11 smaller than the central value of Ref.\ \cite{magnus90}, a discrepancy we attribute to imperfect background subtraction in the previous determination. To obtain the resonance strengths we take a weighted average of B$_{\alpha}$ where more than one measurement is available, excepting the 4.379 MeV state for which our upper limit is preferred. The adopted values are shown in Table\ \ref{table}. The uncertainties given in the table are 1$\sigma$ values, and we specify all upper limits at the 90\% confidence level.

The experimental data on $\Gamma_{\gamma}$ for states in $^{19}$Ne are sparse. Of the six states considered here, measurements are available only for the 4.033 MeV state. The value adopted for this state \cite{hackman00} is the result of a combined analysis of Coulomb excitation and Doppler shift attenuation \cite{davidson73} data using shell-model calculations of the relative strengths of E2 and M1 transitions. In some cases, widths of the analog state from the mirror nucleus $^{19}$F have been measured, and we adopt these under the assumption that  $\Gamma_{\gamma}(^{19}$Ne)~=~$\Gamma_{\gamma}(^{19}$F). Such measurements are available for the 4.549 \cite{tilley95}, 4.600 \cite{kiss82}, and 4.712 MeV states \cite{tilley95}. For the 4.379  and 5.092 MeV states, measurements in neither nucleus are available, and we adopt the results of shell-model calculations \cite{brown02}, assigning a 1$\sigma$ uncertainty of 20\% to the calculated widths. The values of $\Gamma_{\gamma}$ and $\Gamma_{\alpha}$, which is calculated as $\Gamma_{\alpha}=\frac{B_{\alpha}}{1-B_{\alpha}} \Gamma_{\gamma}$, are shown in Table\ \ref{table}. For the 4.033 and 4.379 MeV states we calculate upper limits on $\Gamma_{\alpha}$ at the 90\% confidence level using 1.64$\sigma$ upper limits on $\Gamma_{\gamma}$. Earlier compilations of decay widths can be found in Refs.\ \cite{wilmes95,oliveira97}.

\begin{table*}
 \caption{Branching Ratios B$_{\alpha}\equiv\Gamma_{\alpha}/\Gamma$ and Decay Widths. Upper limits are specified at the 90\% confidence level.\label{table}}
 \begin{ruledtabular}
 \begin{tabular}{ccccccccc}
E$_{x}$ (MeV)&J$^{\pi}$&B$_{\alpha}$ (present work)&B$_{\alpha}$ (Ref.\ \cite{magnus90}) & B$_{\alpha}$ (Ref.\ \cite{laird01})&B$_{\alpha}$ (adopted)&$\Gamma_{\gamma}$ (meV) & Ref. &$\Gamma_{\alpha}$ (meV)\\
\hline
4.033 &$\frac{3}{2}^+$&$\leq$ 4.3 $\times10^{-4}$& & &$\leq$ 4.3 $\times10^{-4}$&12 $^{+9}_{-5}$& \cite{hackman00}&$\leq$ 0.011\\
4.379 &$\frac{7}{2}^+$&$\leq$ 3.9 $\times10^{-3}$&0.044 $\pm$ 0.032 & &$\leq$ 3.9 $\times10^{-3}$&458 $\pm$ 92 & \cite{brown02}&$\leq$ 2.4\\
4.549 &($\frac{1}{2},\frac{3}{2})^-$&  0.16 $\pm$ 0.04 & 0.07 $\pm$ 0.03 & &0.10 $\pm$ 0.02 &39 $^{+34}_{-15}$ & \cite{tilley95}&4.4 $^{+ 4.0}_{- 2.0}$\\
4.600 &($\frac{5}{2}^+$)& 0.32 $\pm$ 0.04& 0.25 $\pm$ 0.04 & $0.32\pm0.03$ & 0.30 $\pm$ 0.02&101 $\pm$ 55 & \cite{kiss82} &43 $\pm$ 24\\
4.712 &($\frac{5}{2}^-$)& 0.85 $\pm$ 0.04 & 0.82 $\pm$ 0.15 & & 0.85 $\pm$ 0.04&43 $\pm$ 8& \cite{tilley95}&230 $\pm$ 80\\\
5.092 &$\frac{5}{2}^+$& 0.90 $\pm$ 0.06 & 0.90 $\pm$ 0.09 & & 0.90 $\pm$ 0.05&196 $\pm$ 39 & \cite{brown02} &1800 $\pm$ 1000\\
 \end{tabular}
\end{ruledtabular}
 \end{table*} 

\begin{figure}\includegraphics[width=\linewidth]{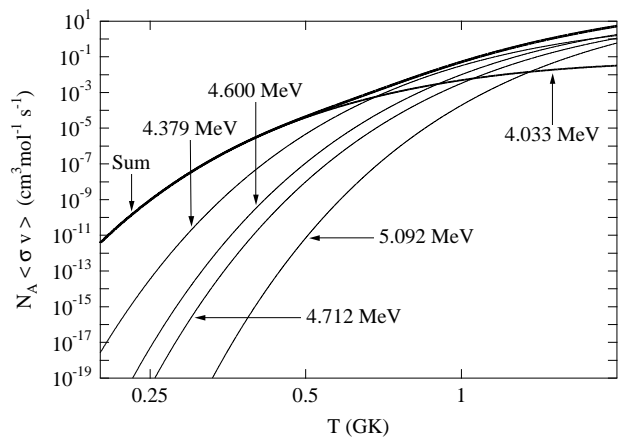} \caption{Product of the Avogadro constant N$_{\mathrm{A}}$ and the thermally averaged rate of the $^{15}$O($\alpha$,$\gamma$)$^{19}$Ne reaction per particle pair. Contributions from all of the important resonances are shown, along with the sum of the resonant and direct capture rates. The contributions of the 4.033 and 4.379 MeV states are 99.73\% confidence level upper limits.} \label{fig3} \end{figure}

We have calculated (see e.g. \cite{rolfs88}) thermally averaged reaction rates due to these 6 resonances. Both the individual rates and the sum of the resonant and direct capture contributions are shown in Fig.\ \ref{fig3}. The contributions of the 4.033 and 4.379 MeV states are calculated using our 99.73\% confidence level upper limits for their $\alpha$-widths, 31 $\mu$eV and 5.6 meV respectively. We do not show the individual contribution of the 4.549 MeV state because it is insignificant by comparison with the other resonances. The direct capture rate was calculated as in Ref.\ \cite{langanke86}, but is significant only below 0.1 GK. Our upper limit on the contribution of the 4.033 MeV state is much larger than all other contributions to the reaction rate for T $\leq$ 0.5 GK. On the contrary, the 4.600 and 4.712 MeV states account for most of the reaction rate at the high temperatures of 1.9 GK found in accreting neutron stars \cite{schatz01}. 

\begin{figure}\includegraphics[width=\linewidth]{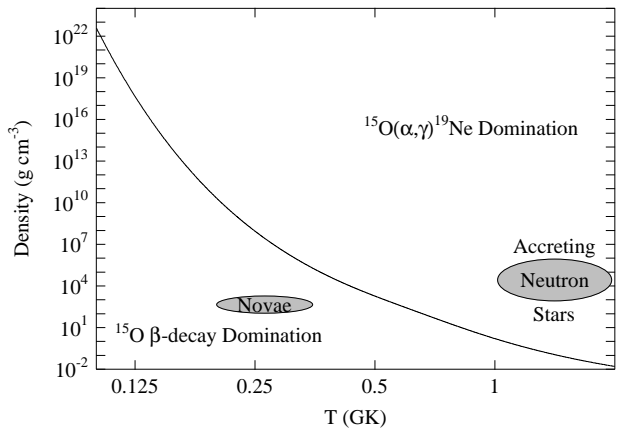} \caption{Density at which the rate of the $^{15}$O($\alpha$,$\gamma$)$^{19}$Ne reaction equals the $\beta^+$ decay rate of $^{15}$O for a He mass fraction of 0.27, the solar value. Below 0.5 GK, the curve represents our 99.73\% confidence level lower limit. The shaded regions indicate the peak temperature and density conditions found in novae and accreting neutron stars.} \label{fig4} \end{figure}

The amount of leakage from the hot CNO cycle via $^{15}$O($\alpha$,$\gamma$)$^{19}$Ne depends on its rate compared to the $\beta^+$ decay rate of $^{15}$O (t$_{1/2}$ = 122 s). In order to calculate the reaction rate in a particular environment one needs to know the local He mass fraction Y. We assume here a Y of 0.27, which is the solar value \cite{anders89} adopted in accreting neutron star models \cite{schatz01}, but is approximately twice the maximum value used in nova models \cite{wanajo99,starrfield00,jose01}. Although the accreted material in novae is usually assumed to be of solar composition, significant mixing with the surface material of the white dwarf occurs prior to the nova outburst \cite{rosner01}, rendering the net Y smaller than that of the accreted matter. For this reason and because we adopt a 99.73\% confidence level upper limit for the contribution of the 4.033 MeV state, the reaction rate we calculate with this Y represents an extreme upper limit for novae. Fig.\ \ref{fig4} shows the boundary in the density-temperature plane at which the rate of the $^{15}$O($\alpha$,$\gamma$)$^{19}$Ne reaction equals the $\beta^+$ decay rate of $^{15}$O for a Y of 0.27. Also shown are shaded regions corresponding to the peak temperatures and densities reached in nova outbursts and accreting neutron stars. A comparable figure appears in Ref.\ \cite{vancraeynest98}.

Typical nova models reach peak temperatures of 0.2~-~0.3 GK, at which time densities are of order 100 g cm$^{-3}$ \cite{starrfield00,jose98,jose01}. Under such conditions, the $\beta^+$ decay rate of $^{15}$O is more than 3 orders of magnitude faster than the rate of the $^{15}$O($\alpha$,$\gamma$)$^{19}$Ne reaction. At temperatures below 0.1 GK in novae, energy is generated by the $pp$ chains and the cold CNO cycle. Only at temperatures above 0.1 GK can the $^{13}$N($p,\gamma)^{14}$O reaction compete equally with $^{13}$N $\beta^+$ decay, initiating the hot CNO cycle. During a nova outburst the temperature exceeds 0.1 GK for about 1000 s \cite{starrfield00,jose98}, while the mean time required to complete one loop of the hot CNO cycle, determined by the $\beta^+$ halflives of $^{14}$O and $^{15}$O and the $^{13}$N($p,\gamma)^{14}$O rate, is at least 300 s. Hence no more than a few cycles can be completed during a nova explosion, and the fractional leakage per cycle through $^{15}$O($\alpha$,$\gamma$)$^{19}$Ne of $<$ 0.001 cannot process a significant amount of CNO material to higher masses. Since the rate of the $^{18}$Ne($\alpha$,$p$)$^{21}$Na reaction appears far too small in novae to compete with $^{18}$Ne $\beta^+$ decay \cite{wiescher99, bradfield-smith99,chen01}, we conclude that appreciable breakout from the hot CNO cycle into the $rp$ process in novae is precluded given our current knowledge of reaction rates and nova physics. This conclusion is consistent with those reached in a recent study of reaction rate variations on nova nucleosynthesis \cite{iliadis02}, which considered a range of $^{15}$O($\alpha$,$\gamma$)$^{19}$Ne rates from 0.002 to 30 times the rate used here.

In summary, we have measured the $\alpha$-decay branching ratios for all of the states in $^{19}$Ne relevant to the astrophysical rate of the $^{15}$O($\alpha$,$\gamma$)$^{19}$Ne reaction, populating them by means of the $p(^{21}$Ne,$t)^{19}$Ne reaction at 43 MeV/u and observing their decays with 100\% geometric efficiency in a magnetic spectrometer. Combining our measurements with prior determinations of the $\gamma$-widths of these states, we have calculated the astrophysical rate of $^{15}$O($\alpha$,$\gamma$)$^{19}$Ne. For the first time, the calculation of the reaction rate at T $\leq$ 0.5 GK is based on direct experimental measurements. On the basis of these calculations and independent determinations of the $^{18}$Ne($\alpha$,$p$)$^{21}$Na reaction rate, we conclude that there can be no appreciable breakout from the hot CNO cycle into the $rp$ process in novae, assuming current models accurately represent their temperature and density conditions.

This work was performed as part of the research program of the {\it Stichting voor Fundamenteel Onderzoek der Materie} with financial support from the {\it Nederlandse Organisatie voor Wetenschappelijk Onderzoek}. KER, RHS, and AHW acknowledge support from a NATO Collaborative Linkage Grant. The ANL Physics Division is supported by the U. S. Department of Energy Nuclear Physics Division under Contract No. W-31-109-Eng38. RES acknowledges support from the US DOE under contract DE-FG02-98ER41086.
\bibliography{prl}

\begin{thebibliography}{35}
\expandafter\ifx\csname natexlab\endcsname\relax\def\natexlab#1{#1}\fi
\expandafter\ifx\csname bibnamefont\endcsname\relax
  \def\bibnamefont#1{#1}\fi
\expandafter\ifx\csname bibfnamefont\endcsname\relax
  \def\bibfnamefont#1{#1}\fi
\expandafter\ifx\csname citenamefont\endcsname\relax
  \def\citenamefont#1{#1}\fi
\expandafter\ifx\csname url\endcsname\relax
  \def\url#1{\texttt{#1}}\fi
\expandafter\ifx\csname urlprefix\endcsname\relax\def\urlprefix{URL }\fi
\providecommand{\bibinfo}[2]{#2}
\providecommand{\eprint}[2][]{\url{#2}}

\bibitem[{\citenamefont{{Gehrz} et~al.}(1998)}]{gehrz98}
\bibinfo{author}{\bibfnamefont{R.~D.} \bibnamefont{{Gehrz}}}
  \bibnamefont{et~al.}, \bibinfo{journal}{Pub. Astron. Soc. Pac.}
  \textbf{\bibinfo{volume}{110}}, \bibinfo{pages}{3} (\bibinfo{year}{1998}).

\bibitem[{\citenamefont{Wallace and Woosley}(1981)}]{wallace81}
\bibinfo{author}{\bibfnamefont{R.~K.} \bibnamefont{Wallace}} \bibnamefont{and}
  \bibinfo{author}{\bibfnamefont{S.~E.} \bibnamefont{Woosley}},
  \bibinfo{journal}{Astrophys. J. Suppl. Ser.} \textbf{\bibinfo{volume}{45}},
  \bibinfo{pages}{389} (\bibinfo{year}{1981}).

\bibitem[{\citenamefont{Schatz et~al.}(2001)}]{schatz01}
\bibinfo{author}{\bibfnamefont{H.}~\bibnamefont{Schatz}} \bibnamefont{et~al.},
  \bibinfo{journal}{Phys. Rev. Lett.} \textbf{\bibinfo{volume}{86}},
  \bibinfo{pages}{3471} (\bibinfo{year}{2001}).

\bibitem[{\citenamefont{Wiescher et~al.}(1999)\citenamefont{Wiescher,
  G\"{o}rres, and Schatz}}]{wiescher99}
\bibinfo{author}{\bibfnamefont{M.}~\bibnamefont{Wiescher}},
  \bibinfo{author}{\bibfnamefont{J.}~\bibnamefont{G\"{o}rres}},
  \bibnamefont{and} \bibinfo{author}{\bibfnamefont{H.}~\bibnamefont{Schatz}},
  \bibinfo{journal}{J. Phys. G} \textbf{\bibinfo{volume}{25}},
  \bibinfo{pages}{R133} (\bibinfo{year}{1999}).

\bibitem[{\citenamefont{Langanke et~al.}(1986)}]{langanke86}
\bibinfo{author}{\bibfnamefont{K.}~\bibnamefont{Langanke}}
  \bibnamefont{et~al.}, \bibinfo{journal}{Astrophys. J.}
  \textbf{\bibinfo{volume}{301}}, \bibinfo{pages}{629} (\bibinfo{year}{1986}).

\bibitem[{\citenamefont{Dufour and Descouvemont}(2000)}]{dufour00}
\bibinfo{author}{\bibfnamefont{M.}~\bibnamefont{Dufour}} \bibnamefont{and}
  \bibinfo{author}{\bibfnamefont{P.}~\bibnamefont{Descouvemont}},
  \bibinfo{journal}{Nucl. Phys.} \textbf{\bibinfo{volume}{A672}},
  \bibinfo{pages}{153} (\bibinfo{year}{2000}).

\bibitem[{\citenamefont{Mao et~al.}(1996)\citenamefont{Mao, Fortune, and
  Lacaze}}]{mao96}
\bibinfo{author}{\bibfnamefont{Z.~Q.} \bibnamefont{Mao}},
  \bibinfo{author}{\bibfnamefont{H.~T.} \bibnamefont{Fortune}},
  \bibnamefont{and} \bibinfo{author}{\bibfnamefont{A.~G.}
  \bibnamefont{Lacaze}}, \bibinfo{journal}{Phys. Rev. C}
  \textbf{\bibinfo{volume}{53}}, \bibinfo{pages}{1197} (\bibinfo{year}{1996}).

\bibitem[{\citenamefont{de~Oliveira et~al.}(1997)}]{oliveira97}
\bibinfo{author}{\bibfnamefont{F.}~\bibnamefont{de~Oliveira}}
  \bibnamefont{et~al.}, \bibinfo{journal}{Phys. Rev. C}
  \textbf{\bibinfo{volume}{55}}, \bibinfo{pages}{3149} (\bibinfo{year}{1997}).

\bibitem[{\citenamefont{Audi and Wapstra}(1995)}]{audi95}
\bibinfo{author}{\bibfnamefont{G.}~\bibnamefont{Audi}} \bibnamefont{and}
  \bibinfo{author}{\bibfnamefont{A.~H.} \bibnamefont{Wapstra}},
  \bibinfo{journal}{Nucl. Phys.} \textbf{\bibinfo{volume}{A595}},
  \bibinfo{pages}{409} (\bibinfo{year}{1995}).

\bibitem[{\citenamefont{Magnus et~al.}(1990)}]{magnus90}
\bibinfo{author}{\bibfnamefont{P.~V.} \bibnamefont{Magnus}}
  \bibnamefont{et~al.}, \bibinfo{journal}{Nucl. Phys.}
  \textbf{\bibinfo{volume}{A506}}, \bibinfo{pages}{332} (\bibinfo{year}{1990}).

\bibitem[{\citenamefont{{Kubono} et~al.}(2002)}]{kubono02}
\bibinfo{author}{\bibfnamefont{S.}~\bibnamefont{{Kubono}}}
  \bibnamefont{et~al.}, \bibinfo{journal}{European Physical Journal A}
  \textbf{\bibinfo{volume}{13}}, \bibinfo{pages}{217} (\bibinfo{year}{2002}).

\bibitem[{\citenamefont{Wiescher}(2002)}]{wiescher02}
\bibinfo{author}{\bibfnamefont{M.}~\bibnamefont{Wiescher}},
  \bibinfo{howpublished}{private communication} (\bibinfo{year}{2002}).

\bibitem[{\citenamefont{Rehm et~al.}(2000)}]{rehm00}
\bibinfo{author}{\bibfnamefont{K.~E.} \bibnamefont{Rehm}} \bibnamefont{et~al.},
  \emph{\bibinfo{title}{Physics Division Annual Report}}
  (\bibinfo{publisher}{Argonne National Laboratory}, \bibinfo{year}{2000}),
  p.~\bibinfo{pages}{6}.

\bibitem[{\citenamefont{van~den Berg}(1995)}]{berg95}
\bibinfo{author}{\bibfnamefont{A.~M.} \bibnamefont{van~den Berg}},
  \bibinfo{journal}{Nucl. Instrum. Methods} \textbf{\bibinfo{volume}{B99}},
  \bibinfo{pages}{637} (\bibinfo{year}{1995}).

\bibitem[{\citenamefont{{Fortune} et~al.}(1978)\citenamefont{{Fortune}, {Nann},
  and {Wildenthal}}}]{fortune78}
\bibinfo{author}{\bibfnamefont{H.~T.} \bibnamefont{{Fortune}}},
  \bibinfo{author}{\bibfnamefont{H.}~\bibnamefont{{Nann}}}, \bibnamefont{and}
  \bibinfo{author}{\bibfnamefont{B.~H.} \bibnamefont{{Wildenthal}}},
  \bibinfo{journal}{\prc} \textbf{\bibinfo{volume}{18}}, \bibinfo{pages}{1563}
  (\bibinfo{year}{1978}).

\bibitem[{\citenamefont{W\"{o}rtche}(2001)}]{woertche01}
\bibinfo{author}{\bibfnamefont{H.~J.} \bibnamefont{W\"{o}rtche}},
  \bibinfo{journal}{Nucl. Phys.} \textbf{\bibinfo{volume}{A687}},
  \bibinfo{pages}{321c} (\bibinfo{year}{2001}).

\bibitem[{\citenamefont{Leegte et~al.}(1992)}]{leegte92}
\bibinfo{author}{\bibfnamefont{H.~K.~W.} \bibnamefont{Leegte}}
  \bibnamefont{et~al.}, \bibinfo{journal}{Nucl. Instrum. Methods}
  \textbf{\bibinfo{volume}{A313}}, \bibinfo{pages}{260} (\bibinfo{year}{1992}).

\bibitem[{\citenamefont{Tilley et~al.}(1995)}]{tilley95}
\bibinfo{author}{\bibfnamefont{D.~R.} \bibnamefont{Tilley}}
  \bibnamefont{et~al.}, \bibinfo{journal}{Nucl. Phys.}
  \textbf{\bibinfo{volume}{A595}}, \bibinfo{pages}{1} (\bibinfo{year}{1995}).

\bibitem[{\citenamefont{Laird et~al.}(2001)}]{laird01}
\bibinfo{author}{\bibfnamefont{A.~M.} \bibnamefont{Laird}}
  \bibnamefont{et~al.}, \bibinfo{journal}{Nucl. Phys.}
  \textbf{\bibinfo{volume}{A688}}, \bibinfo{pages}{134c}
  (\bibinfo{year}{2001}).

\bibitem[{\citenamefont{Hackman et~al.}(2000)}]{hackman00}
\bibinfo{author}{\bibfnamefont{G.}~\bibnamefont{Hackman}} \bibnamefont{et~al.},
  \bibinfo{journal}{Phys. Rev. C} \textbf{\bibinfo{volume}{61}},
  \bibinfo{pages}{052801} (\bibinfo{year}{2000}).

\bibitem[{\citenamefont{Davidson and Roush}(1973)}]{davidson73}
\bibinfo{author}{\bibfnamefont{J.~M.} \bibnamefont{Davidson}} \bibnamefont{and}
  \bibinfo{author}{\bibfnamefont{M.~L.} \bibnamefont{Roush}},
  \bibinfo{journal}{Nucl. Phys.} \textbf{\bibinfo{volume}{A213}},
  \bibinfo{pages}{332} (\bibinfo{year}{1973}).

\bibitem[{\citenamefont{Kiss et~al.}(1982)}]{kiss82}
\bibinfo{author}{\bibfnamefont{{\'{A}}.~Z.} \bibnamefont{Kiss}}
  \bibnamefont{et~al.}, \bibinfo{journal}{Nucl. Instrum. Methods}
  \textbf{\bibinfo{volume}{203}}, \bibinfo{pages}{107} (\bibinfo{year}{1982}).

\bibitem[{\citenamefont{Brown}(2002)}]{brown02}
\bibinfo{author}{\bibfnamefont{B.~A.} \bibnamefont{Brown}},
  \bibinfo{howpublished}{private communication} (\bibinfo{year}{2002}).

\bibitem[{\citenamefont{{Wilmes} et~al.}(1995)}]{wilmes95}
\bibinfo{author}{\bibfnamefont{S.}~\bibnamefont{{Wilmes}}}
  \bibnamefont{et~al.}, \bibinfo{journal}{\prc} \textbf{\bibinfo{volume}{52}},
  \bibinfo{pages}{2823} (\bibinfo{year}{1995}).

\bibitem[{\citenamefont{Rolfs and Rodney}(1988)}]{rolfs88}
\bibinfo{author}{\bibfnamefont{C.~E.} \bibnamefont{Rolfs}} \bibnamefont{and}
  \bibinfo{author}{\bibfnamefont{W.~S.} \bibnamefont{Rodney}},
  \emph{\bibinfo{title}{Cauldrons in the Cosmos}} (\bibinfo{publisher}{The
  University of Chicago Press}, \bibinfo{address}{Chicago},
  \bibinfo{year}{1988}).

\bibitem[{\citenamefont{{Anders} and {Grevesse}}(1989)}]{anders89}
\bibinfo{author}{\bibfnamefont{E.}~\bibnamefont{{Anders}}} \bibnamefont{and}
  \bibinfo{author}{\bibfnamefont{N.}~\bibnamefont{{Grevesse}}},
  \bibinfo{journal}{Geochim. Cosmochim. Acta} \textbf{\bibinfo{volume}{53}},
  \bibinfo{pages}{197} (\bibinfo{year}{1989}).

\bibitem[{\citenamefont{Wanajo et~al.}(1999)\citenamefont{Wanajo, Hashimoto,
  and Nomoto}}]{wanajo99}
\bibinfo{author}{\bibfnamefont{S.}~\bibnamefont{Wanajo}},
  \bibinfo{author}{\bibfnamefont{M.-A.} \bibnamefont{Hashimoto}},
  \bibnamefont{and} \bibinfo{author}{\bibfnamefont{K.}~\bibnamefont{Nomoto}},
  \bibinfo{journal}{Astrophys. J.} \textbf{\bibinfo{volume}{523}},
  \bibinfo{pages}{409} (\bibinfo{year}{1999}).

\bibitem[{\citenamefont{Starrfield et~al.}(2000)}]{starrfield00}
\bibinfo{author}{\bibfnamefont{S.}~\bibnamefont{Starrfield}}
  \bibnamefont{et~al.}, \bibinfo{journal}{Astrophys. J. Suppl. Ser.}
  \textbf{\bibinfo{volume}{127}}, \bibinfo{pages}{485} (\bibinfo{year}{2000}).

\bibitem[{\citenamefont{{Jos{\' e}} et~al.}(2001)\citenamefont{{Jos{\' e}},
  {Coc}, and {Hernanz}}}]{jose01}
\bibinfo{author}{\bibfnamefont{J.}~\bibnamefont{{Jos{\' e}}}},
  \bibinfo{author}{\bibfnamefont{A.}~\bibnamefont{{Coc}}}, \bibnamefont{and}
  \bibinfo{author}{\bibfnamefont{M.}~\bibnamefont{{Hernanz}}},
  \bibinfo{journal}{\apj} \textbf{\bibinfo{volume}{560}}, \bibinfo{pages}{897}
  (\bibinfo{year}{2001}).

\bibitem[{\citenamefont{{Rosner} et~al.}(2001)}]{rosner01}
\bibinfo{author}{\bibfnamefont{R.}~\bibnamefont{{Rosner}}}
  \bibnamefont{et~al.}, \bibinfo{journal}{Astrophys. J.}
  \textbf{\bibinfo{volume}{562}}, \bibinfo{pages}{L177} (\bibinfo{year}{2001}).

\bibitem[{\citenamefont{{Vancraeynest} et~al.}(1998)}]{vancraeynest98}
\bibinfo{author}{\bibfnamefont{G.}~\bibnamefont{{Vancraeynest}}}
  \bibnamefont{et~al.}, \bibinfo{journal}{\prc} \textbf{\bibinfo{volume}{57}},
  \bibinfo{pages}{2711} (\bibinfo{year}{1998}).

\bibitem[{\citenamefont{Jos\'{e} and Hernanz}(1998)}]{jose98}
\bibinfo{author}{\bibfnamefont{J.}~\bibnamefont{Jos\'{e}}} \bibnamefont{and}
  \bibinfo{author}{\bibfnamefont{M.}~\bibnamefont{Hernanz}},
  \bibinfo{journal}{Astrophys. J.} \textbf{\bibinfo{volume}{494}},
  \bibinfo{pages}{680} (\bibinfo{year}{1998}).

\bibitem[{\citenamefont{{Bradfield-Smith} et~al.}(1999)}]{bradfield-smith99}
\bibinfo{author}{\bibfnamefont{W.}~\bibnamefont{{Bradfield-Smith}}}
  \bibnamefont{et~al.}, \bibinfo{journal}{Phys. Rev. C}
  \textbf{\bibinfo{volume}{59}}, \bibinfo{pages}{3402} (\bibinfo{year}{1999}).

\bibitem[{\citenamefont{{Chen} et~al.}(2001)}]{chen01}
\bibinfo{author}{\bibfnamefont{A.~A.} \bibnamefont{{Chen}}}
  \bibnamefont{et~al.}, \bibinfo{journal}{Phys. Rev. C}
  \textbf{\bibinfo{volume}{63}}, \bibinfo{pages}{65807} (\bibinfo{year}{2001}).

\bibitem[{\citenamefont{Iliadis et~al.}(2002)}]{iliadis02}
\bibinfo{author}{\bibfnamefont{C.}~\bibnamefont{Iliadis}} \bibnamefont{et~al.},
  \bibinfo{journal}{Astrophys. J. Suppl. Ser.} \textbf{\bibinfo{volume}{142}},
  \bibinfo{pages}{105} (\bibinfo{year}{2002}).

\end{thebibliography}
\end{document}